# Neutrino driven explosions aided by axion cooling in multidimensional simulations of core-collapse supernovae

Aurore Betranhandy and Evan O'Connor
*The Oskar Klein Centre, Department of Astronomy, Stockholm University,
AlbaNova, SE-106 91 Stockholm, Sweden*

In this study, we present the first multidimensional core-collapse supernovae (CCSNe) simulations including QCD axions in order to assess the impact on the CCSN explosion mechanism. We include axions in our simulations through the nucleon-nucleon bremsstrahlung emission channel and as a pure energy-sink term under the assumption that the axions free-stream after being emitted. We perform both spherically symmetric (1D) and axisymmetric (2D) simulations. In 1D, we utilize a parametrized heating scheme to achieve explosions, whereas in 2D we self-consistently realize explosions through the neutrino heating mechanism. Our 2D results for a 20 $M_\odot$ progenitor show an impact of the axion emission on the shock behavior and the explosion time when considering values of the PecceiQuinn energy scale $f_a \leq 2 \times 10^8$ GeV. The strong cooling due to the axion emission accelerates the contraction of the core and leads to more efficient neutrino heating and earlier explosions. For the axion emission formalism utilized, the values of $f_a$ that impact the explosion are close to, but in tension with current limits based on the neutrinos detected from SN 1987A. However, given the nonlinear behavior of the emission and the multidimensional nature of CCSNe, we suggest that a self-consistent, multidimensional approach to simulating CCSNe, including any late time accretion and cooling, is needed to fully explore the axion bounds from supernovae and the impact on the CCSN explosion mechanism.

DOI: 10.1103/PhysRevD.106.063019

## I. INTRODUCTION

Core-Collapse Supernovae (CCSNe), as the final stage of the life of massive stars, are a key piece of the puzzle for our understanding of stellar evolution, and are an important source of new elements through the nucleosynthesis happening during the event. They also are the birthplace of neutron stars and black holes, therefore they play a major role in setting the population of compact object mergers in the Universe. They are such an extreme environment, that they also provide a fantastic cosmic laboratory for studying and testing particle physics. This includes neutrinos and axions, the latter of which are the focus of this study. While CCSNe have been observed for a long time, it is mainly through electromagnetic radiation. This channel gives us precious information about the explosion energy, nucleosynthesis, and the outer parts of the progenitor, but fails to inform us about the collapse and the initial explosion mechanism. This is because when the explosion is launched, the outer parts of the star are still opaque to photons. The other observational channels that could give us information about the explosion mechanism are neutrinos and gravitational waves (GW) [1–5]. However, these channels are only detectable in the unlikely event of a galactic CCSN. This makes simulations critical in developing our understanding of the core collapse. From these simulations, the general evolution of a CCSN is fairly well understood [6,7]. When the star collapses and reaches nuclear densities, the core stiffens to a point where the collapse is halted and initially reversed, the so-called bounce. The pressure waves associated with this bounce travel through the infalling matter, and eventually become supersonic, creating a shock. While propagating out, this shock loses energy through the dissociation of heavy nuclei and emission of neutrinos. This energy loss will cause the shock to stall at 150–200 km from the core, succumbing to the ram pressure of the still infalling matter above the shock. It then needs to be revived in order to create the final explosion. Different revival mechanisms have been explored through the years [6,8–12] but two main theories are well recognized as viable, the neutrino driven explosion mechanism and the magneto-rotational explosion mechanism. The neutrino-driven explosion mechanism relies on the transfer of energy from the cooling protoneutron star (PNS) to the material behind the shock via neutrino heating, helped by hydrodynamic turbulence.

This mechanism is robust and has been studied extensively in the last years, up to the point where we now obtain self-consistent explosions in 3D. In this paper, we focus on this mechanism, ignoring the potential role of magnetic field and rotation in our simulations.

While it is possible to now obtain self-consistent explosions in 3D, the explosion energy is regularly under the $\sim 10^{51}$ erg secular threshold, and only a few explosions manage to reach this value [13]. This lack of explosion energy is potentially due to the approximations used in our simulations or missing physics. One piece of missing physics could be the disregard of axion physics. Axions are pseudo-Nambu-Goldstone bosons theorized to preserve $CP$-symmetry in QCD [14]. They have not been observed and are therefore purely theoretical but remain one of the main candidates to resolve the "strong $CP$ problem". While axions have been previously researched in the scope of CCSNe, the number of actual simulations including these particles is low [15–17] and no multidimensional exploration has been done. While axions are purely theoretical, there are many constraints of their coupling constant with the matter. Some of these constraints have been placed following the neutrino detection from SN 1987A [18–22]. However these studies were made following simple prescriptions for the supernovae explosion. Axion emission in a CCSN will impact the cooling of the PNS by acting as an alternative sink term for the energy. It is worth investigating, with self-consistent multidimensional simulations, whether and in what way, the axion emission will impact the neutrino production and therefore the energy that can be transported to the shock to impact its revival. In this paper, we focus on axion production via nucleon-nucleon bremsstrahlung using the One Pion Exchange (OPE) formalism [18,23,24]. However, recent studies have been done using other production channels such as pion interactions [15,25], the Primakoff process and photon coalescence [17], and improved interactions using a different formalism of OPE [15,23]. Most of these studies highlight the importance of including axions in CCSNe simulations with Mori *et al.* [17] even obtaining an explosion in 1D when including Axion-like particles (ALPs) with masses $m_a = 40$–$400$ MeV produced through Primakoff process and photon coalescence.

In this paper, we present a set of simulations that includes the impact of cooling via axion emission through nucleon-nucleon bremsstrahlung using an OPE formalism [18,23]. Our main goal is to explore, for the first time with self-consistent multidimensional neutrino-radiation hydrodynamic simulations, the impact of various axion-matter coupling strengths and to assess for what values axions can have an impact on the explosion mechanism itself. If there is an impact, does this help or hinder the development of an explosion? We do not strictly limit ourselves to values of the coupling constant that obey constraints set by SN 1987A, though we are close. This is for two reasons:

1. Constraints set by observations of neutrinos from SN 1987A rarely follow from fully self-consistent simulations of CCSNe including said axions. Indeed, we arguably do not have a full picture of the CCSNe physics even when setting the potential role of axions aside. 2. We want to give ourselves some leeway to fully explore the impact of cooling via axions without being restricted. This being said, SN 1987A does indeed place limits on any potential axion emission, particularly from the duration of the signal. We return to this point in the discussion. We begin by presenting our simulation code FLASH as well as the setup of our simulations in Sec. II A. We then present the method of including axion emission through the nucleon-nucleon bremsstrahlung using the OPE formalism in Sec. II B. In Sec. III, we present our simulations beginning with a reproduction of a former study [16] and an exploration of the impact of different formalisms. Then we present the results of systematic simulations on a well tested progenitor in spherical symmetry (1D) and then axisymmetry (2D) while varying the axion-matter coupling strength. We end the paper with a discussion on potential limitations in our study in Sec. IV before concluding in Sec. V.

## II. METHOD

### A. FLASH and setup

For all the simulations in this paper, we use the FLASH open-source framework for multidimensional hydrodynamic simulations [26] outfitted to simulate CCSNe using a multidimensional, grid-based, neutrino transport [27,28]. For this work, FLASH will be used assuming spherical symmetry (1D) and assuming axisymmetry (2D). In 1D, our domain extends to a radius of $10^9$ cm with a finest grid zone of $\sim 203$ m. Our 2D simulations assume an axisymmetric geometry and use cylindrical coordinates that extend to a radius of $10^9$ cm and cover from $-10^9$ cm to $+10^9$ cm along the polar axis. Using 9 levels of FLASH's adaptive mesh refinement, we achieve a finest grid spacing of $\sim 465$ m in the protoneutron star core and maintain an effective angular resolution of 0.8 degrees outside of 80 km. For our 2D simulations there is no initial rotation and magnetic fields are ignored.

In our simulations, gravity is treated with a modified Newtonian potential (case A from Ref. [29]). We evolve the neutrino radiation fields via a M1-scheme with an analytic closure [28,30–32] and assume 3 neutrino species ($\nu_e$, $\bar{\nu}_e$, and $\nu_x$, where $\nu_x$ is a characteristic heavy-lepton neutrino encompassing the four heavy-lepton neutrino species $\nu_\mu$, $\bar{\nu}_\mu$, $\nu_\tau$, and $\bar{\nu}_\tau$). The neutrino interactions are computed using the NuLib software [31] following [33–35] and include elastic neutrino-nucleon and neutrino-nucleus scattering, charged-current emission and absorption of electron neutrinos and antineutrinos, thermal emission of heavy-lepton neutrino pairs, and inelastic scattering of neutrinos on electrons. Unless otherwise stated, we utilize the baseline SRO equation of state (EOS) from Ref. [36] which is the same baseline EOS used in [1].

The axion emission and the subsequent impact on the evolution of the internal energy is implemented as a pure energy sink term within FLASH's sourceTerms unit. During each time step, for each grid zone, the energy emitted in axions is determined (see the next section for details of the axion emissivity). This is then directly taken out of the matter internal energy and tracked. This has the consequence of cooling the dense protoneutron star core where the bulk of the axion emission is occurring.

In our study we use two progenitors. First, in order to verify our implementation and compare to earlier results from [16], we used the 18 $M_\odot$ progenitor models from [37]. For the remainder of our study we utilize a 20 $M_\odot$ progenitor from [38] to systematically test the axion-matter coupling constant first in 1D and then in 2D. This model has been studied extensively in the last years in 1D, 2D, and 3D simulations [1,28,39–43].

## B. Nucleon-nucleon bremsstrahlung axion emission formalism

In this work, we chose the nucleon-nucleon bremsstrahlung axion emission ($N + N \rightleftarrows N^* + N^* + a$) formalism based on the One Pion Exchange (OPE) formalism first defined in [18] modified following [23] with an additional correction term coming from [16,21]. The emission (in erg g$^{-1}$ s$^{-1}$) is given as

$$Q_a^{(1)} = 64 \frac{c}{\rho(\hbar c)^4} \left(\frac{f}{m_\pi}\right)^4 m_N^{2.5} T^{6.5} \left[\left(1 - \frac{\xi}{3}\right) g_{an}^2 I\left(\frac{\mu_n}{T}, \frac{\mu_n}{T}\right)\right.$$
$$+ \left(1 - \frac{\xi}{3}\right) g_{ap}^2 I\left(\frac{\mu_p}{T}, \frac{\mu_p}{T}\right)$$
$$+ \frac{4(3 + 2\xi)}{9} \left(\frac{g_{an}^2 + g_{ap}^2}{2}\right) I\left(\frac{\mu_n}{T}, \frac{\mu_p}{T}\right)$$
$$\left.+ \frac{8(3 + 2\xi)}{9} \left(\frac{g_{an} + g_{ap}}{2}\right)^2 I\left(\frac{\mu_n}{T}, \frac{\mu_p}{T}\right)\right], \quad (1)$$

$$Q_a = Q_a^{(1)} \min\left[\frac{\Gamma_\sigma^{\max}}{\Gamma_\sigma}\right], \quad (2)$$

$$\Gamma_\sigma = 10 \text{ MeV} \left(\frac{m_N}{938 \text{ MeV}}\right)^2 \rho_{14} T_{\text{MeV}}^{0.5}, \quad (3)$$

where $m_N = 939.56$ MeV and $m_\pi = 139.57$ MeV are the nucleon and pion mass, respectively, $c$ is the speed of light, $\hbar$ is the reduced Planck constant, $T$ is the temperature, $\rho$ is the density, $f$ is the pion-nucleon coupling constant taken as $f = 1$, $g_{aN}$ is the coupling constant between the axion and the nucleon $N$, $\xi$ is a measure of the nucleon degeneracy with $\xi = 0$ for degenerate matter and $\xi = 1.3078$ for nondegenerate matter, and $I$ is a dimensionless function defined in [18] depending on $\mu_N$ the nucleon chemical potentials and the temperature. The emissivity is then corrected using the term defined in Eq. (2) depending on $\Gamma_\sigma$, the effective spin fluctuation rate, defined in Eq. (3) with $\rho_{14}$ the density divided by $10^{14}$ g cm$^{-3}$ and $T_{\text{MeV}}$ the temperature in MeV. The nucleon spin fluctuation rate, $\Gamma_\sigma$, is an important factor in the axion emission and is expected to saturate depending on the average interaction of a nucleon in the nuclear medium. We take the maximum value as $\Gamma_\sigma^{\max} = 60$ MeV. We note that while we use the simplified OPE formalism from [23], we do not include the other improvements developed there: corrections due to finite pion mass, two pion exchange, effective nucleon mass, and multiple scatterings. We note that these corrections have the impact of lowering the overall emissivity of axions by roughly a factor of 10. Furthermore, it is shown in [23] that this factor of 10 is fairly consistent over different PNS structures during the entire CCSN evolution. For our purposes, where we want to test the impact of the axion emission on the explosion mechanism, this amounts to a mere rescaling of the coupling constant.

We invoke the assumption that axions are in the free-streaming regime following previous studies [16,23]. This allows us to consider only the emissivity in the $N + N \rightleftarrows N^* + N^* + a$ interaction. We then use this emissivity as a pure source of energy for the axions, and therefore a pure sink of internal matter energy. We update the internal matter energy in an operator-split fashion, $E_{\text{int}}^{n+1} = E_{\text{int}}^n - \alpha \Delta t Q_a$, with $\Delta t$ taken as the numerical timestep determined by the radiation hydrodynamics and $\alpha = \exp(\phi_{\text{eff}}/c^2)$ is the lapse factor from our effective gravitational potential. The free-streaming assumption and its limitations will be discussed further in Sec. IV.

In this paper we will often refer to the Peccei-Quinn scale $f_a$ as a proxy for the coupling constant between axions and nucleons. The relationship between these two is

$$g_{aN} = C_N \frac{m_N}{f_a}, \quad (4)$$

where $m_N = 939.56$ MeV is the nucleon mass taken to be equal for protons and neutrons and the $C_N$ are model-dependant constants taken from the hadronic axion model KSVZ as $C_p = -0.47$ and $C_n = -0.02$ [23,44]. For reference, Table I shows the different coupling constant

TABLE I. Reference table for $f_a$ and $g_{aN}$ values used in this paper.

| $f_a$ (GeV) | $\|g_{an}\|$ | $\|g_{ap}\|$ |
|---|---|---|
| $5 \times 10^7$ | $3.758 \times 10^{-10}$ | $8.832 \times 10^{-9}$ |
| $1 \times 10^8$ | $1.879 \times 10^{-10}$ | $4.416 \times 10^{-9}$ |
| $1.5 \times 10^8$ | $1.252 \times 10^{-10}$ | $2.939 \times 10^{-9}$ |
| $2 \times 10^8$ | $9.396 \times 10^{-11}$ | $2.208 \times 10^{-9}$ |
| $3 \times 10^8$ | $6.264 \times 10^{-11}$ | $1.472 \times 10^{-9}$ |
| $4 \times 10^8$ | $4.698 \times 10^{-11}$ | $1.104 \times 10^{-9}$ |
| $5 \times 10^8$ | $3.758 \times 10^{-11}$ | $8.832 \times 10^{-10}$ |

values depending on $f_a$ for specific values of $f_a$ used in our simulations.

## III. RESULTS

In this section, we will present the results of our different simulations. We separate the results first by dimension, exploring spherically symmetric (1D) simulations with artificial heating to trigger an explosion and then with axisymmetric (2D) simulations to explore the impact on the explosion mechanism in self-consistent neutrino-driven explosions. For the spherically symmetric simulations, in Sec. III A 1, we attempt a comparison between our implementation and the results of the study from Fischer *et al.* (2016) [16]. We use this comparison to test our implementation and to highlight some of the different formalisms available for the axion emission via nucleon nucleon bremsstrahlung. We then explore the impact of the coupling constant on the explosion mechanism in 1D in Sec. III A 2. This is done using a 20 $M_\odot$ progenitor widely used in the field [1,40,42,43]. Based on our 1D simulations, in Sec. III B, we select several values of the PecceiQuinn scale, $f_a$, and extend our study to axisymmetric 2D simulations to study the impact of the axion emission in a self-consistent explosion.

### A. 1D results

In order to explore how the axion-nucleon coupling impacts the core-collapse explosion mechanism, we begin with systematic tests on 1D models. We divide these results into two different parts, first a comparative test of our implementation with a previous study [16], and second, a systematic study of the impact of axion emission while varying the Peccei-Quinn energy scale value, $f_a$.

#### 1. 18 $M_\odot$

First we explore the axion-matter coupling for an 18 $M_\odot$, solar metallicity progenitor (from [37]) in order to reproduce the results from Fischer *et al.* [16]. As with Fischer *et al.*, we use the DD2 EOS [45,46] and trigger an *ad hoc* explosion by enhancing the neutrino interactions in the low density region behind the shock where the neutrino heating occurs. Following [47], this region is defined as the region with an entropy $s > 6k_B$/baryon and a density $\rho < 10^{10}$ g cm$^{-3}$. In this region both the emissivity and absorption opacity are enhanced with a multiplying factor of 5–7 [47] for the electron neutrinos and antineutrinos. This multiplication factor increases artificially, and predominantly, the heating in the gain region, helping the shock to expand and create an explosion. We choose a value of 6 for our work here which reproduces the shock evolution seen in Ref. [16].

With this setup we perform six simulations varying the axion-matter couplings. First, we perform a simulation without axion-matter coupling, referred to as our reference.

We perform two simulations to reproduce a selection of the results in Fischer *et al.*, particularly those simulations with $g_{ap} = 9 \times 10^{-10}$ and $g_{an} = 0$. We note this roughly corresponds to a value of $f_a \sim 5 \times 10^8$ GeV. These two simulations correspond to choosing $\xi = 0$ (fully degenerate nucleons) and $\xi = 1$ (nondegenerate nucleons) and we utilize the same axion emissivity used in Fischer *et al.* [16] which ultimately comes from Brinkmann & Turner [18] (B&T). Due to an earlier error related to the sign of a crossing term in the matrix kernel from Brinkmann & Turner (see the discussion in [23]), the emissivities used in Fischer *et al.* are different than the emissivity given in Eq. (1). The final three simulations in this comparison use directly the axion emissivity given in Eqs. (1)–(2), assuming $g_{ap} = 9 \times 10^{-10}$ and $g_{an} = 0$ and three different degeneracy factors $\xi = 0$ (fully degenerate nucleons), $\xi = 1.3078$ (nondegenerate nucleons), and a variable $\xi$ where we choose either 0 or 1.3078 in individual computational zones depending on an rough estimate of the nucleon chemical potential [i.e., $\xi = 0$ if $(3\pi^2 n_N)^{2/3}/(2m_N T) > 1$, $\xi = 1.3078$ otherwise]. For all of the simulations outside of this section, we use this latter formalism from [23]. Though we note, as mentioned in Sec. II, we do not include the other improvements developed there: corrections due to finite pion mass, two pion exchange, effective nucleon mass, multiple scatterings, and arbitrary degeneracy.

Given this choice of $f_a$ for these simulations, the influence of the axion emission on the results in Fischer *et al.* was most prominent after the first second and barely influenced the explosion dynamics shortly after bounce. We see the same impact here. In Fig. 1, we present the shock radius evolution for each of the six simulations described above. For all simulations, the shock initially follows the same behavior, with a first expansion up to $\sim$350 km in the first $\sim$150 ms postbounce. The shock then stalls and begins to recede down, reaching $\sim$200 km at $\sim$200 ms postbounce. A secondary shock then takes the lead and drives the final shock expansion. From $\sim$ 300 ms postbounce, we see a slight divergence of the simulations from the reference simulation with a difference of $\lesssim$5% in the shock radius value. While small due to the small value of the coupling constant $g_{ap}$ (or equivalently the large value of $f_a$), the hierarchy of shock radius evolutions away from the reference simulation is consistent with the observed hierarchy of the axion emission, which we explore below. This foreshadows the results in the remaining sections on the axion emission impact on the explosion mechanism.

For the total neutrino and axion luminosity evolution with time, we refer to Fig. 2. The solid lines show the total ($\nu_e + \bar{\nu}_e + 4\nu_x$) neutrino luminosity and the dashed lines show the total axion luminosity. As with the shock radius, the total neutrino luminosity in all six variants of this simulation is essentially the same up to $\sim$1 s after bounce. For all simulations, the total neutrino luminosity first shows a peak corresponding to the electron neutrino burst following core bounce. It then plateaus at $\sim 2 \times 10^{53}$ erg s$^{-1}$ up to $\sim 200$ ms postbounce where it shows a small luminosity burst corresponding to the accretion onto the PNS coming from the first shock receding. After the plateau phase, when the explosion sets in, the total neutrino luminosity begins to drop as no matter is now accreting onto the PNS. The remaining neutrino luminosity comes from the cooling PNS. Around 1 s after bounce, the total neutrino luminosity begins to diverge between the different simulations. The lowest neutrino luminosity arises from the simulation using the B&T formalism in the degenerate case ($\xi = 0$), followed by the B&T nondegenerate case ($\xi = 1$). The next two lowest neutrino luminosities arise from the simulations using Eqs. (1)–(2) (the Carenza formalism; [23]). These show an inverted hierarchy for the neutrino luminosity as compared to the B&T formalism. The simulation with $\xi = 0$ has a higher total neutrino luminosity than the simulation with $\xi = 1.3078$. The simulation with a varying $\xi$ has an evolution very similar to the simulation with $\xi = 0$. All variants have a lower total neutrino luminosity at this stage when compared to the reference simulation. This is due to the emission of some of the thermal energy as axions. As detailed below, the hierarchy for the magnitude of the peak axion luminosity follows the inverse order of the neutrino luminosity cooling described above. Generally speaking, in the cooling phase, the more energy that is emitted in axions the less energy is emitted in neutrinos as both are ultimately sourced from the same pool of thermal energy.

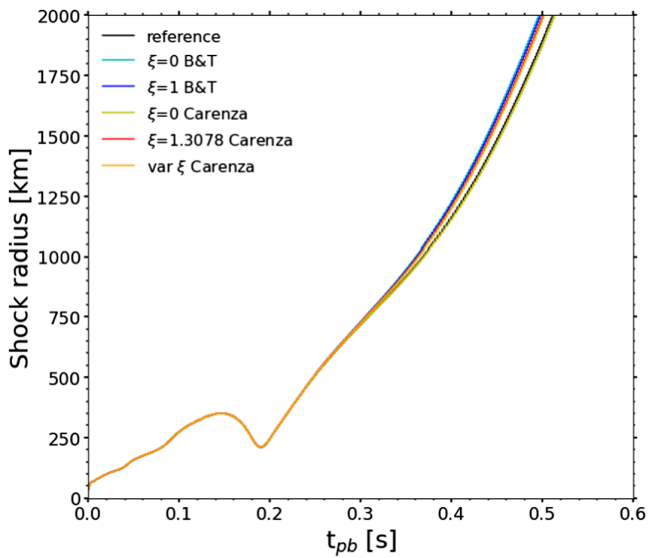

FIG. 1. Shock radius versus the time for a 1D simulation with an 18 $M_\odot$ progenitor in comparison to the results of Fischer *et al.* (2016) [16] (see their Fig. 3). Following Fischer *et al.*, we use artificial neutrino heating to trigger an explosion. All simulations have been run with $g_{ap} = 9 \times 10^{-10}$ and $g_{an} = 0$ or a Peccei-Quinn scale $f_a \sim 5 \times 10^8$ GeV. The different colors correspond to different axion-emission formalisms, although they have little impact on the early shock evolution shown here. The cyan and blue lines result from the formalism of Brinkmann *et al.* [18] (B&T) modified following Eq. (2) and used to match the results of Fischer *et al.* [16], with $\xi$, the degeneracy scale, equal to 0 and 1, respectively. The yellow, red, and orange lines follow the formalism of Carenza *et al.* [23], modified following Eq. (2), with $\xi$ equal to 0, 1.3078, and varying depending on the degeneracy (see text), respectively. The black curve shows the shock radius for the reference simulation without axion emission.

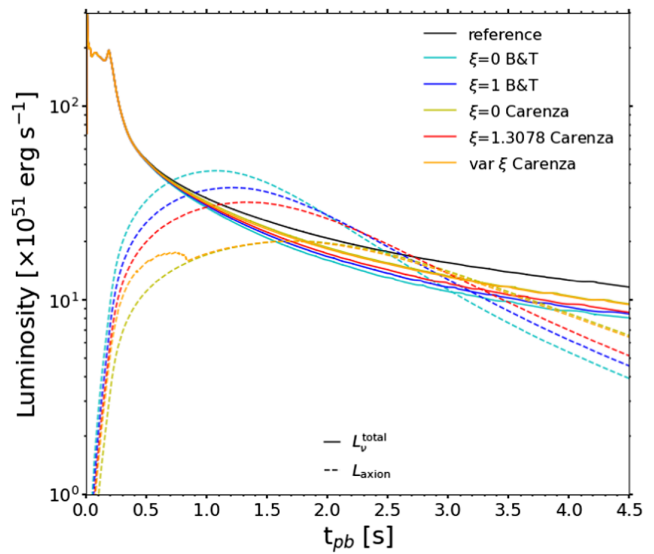

FIG. 2. Neutrino and axion luminosities versus time. The different colors correspond to different axion emission formalisms while the black curve, labeled reference, shows a simulation without axion emission. The cyan and blue lines are following the formalism of Brinkmann *et al.* [18] (B&T) with $\xi$, the degeneracy scale, equal to 0 and 1, respectively, modified following Eq. (2). The yellow, red and orange lines follow the formalism of Carenza *et al.* [23], modified following Eq. (2), with $\xi$ equal to 0, 1.3078, and varying depending on the degeneracy, respectively.

All of the simulations with a fixed $\xi$ have axion luminosities that smoothly evolve through time. There is a sharp increase in the first hundreds of milliseconds, followed by a peak and a slow decrease. The time when the axion emission peaks and the changing rate of the subsequent axion emission varies with the formalism (and choice of $\xi$). Simulations with a higher axion luminosity have a peak occurring earlier, followed by a sharper decrease when compared to a run with a lower overall axion emissivity. This is explained by the different PNSs cooling at different rates due to the continuous loss of energy through the axion emission. We explore this aspect in more detail with the systematic simulations presented in the following section.

Among the different formalisms, the simulations following the B&T formalism have an earlier and higher axion peak luminosity compared to the others. The highest peak axion luminosity is obtained when we assume $\xi = 0$, followed by when we assume $\xi = 1$. These two axion luminosities, in cyan and blue, respectively, should be compared with Fig. 6(a) and Table II of Fischer *et al.* [16]. For our (Fischer *et al.*) $\xi = 0$ simulations we obtain a peak axion luminosity of $\sim 46 \times 10^{51}$ erg s$^{-1}$ ($\sim 37 \times 10^{51}$ erg s$^{-1}$) at a post bounce time of 1.1 s (1.5 s), while for the $\xi = 1$ simulations, we obtain a peak axion luminosity of $\sim 38 \times 10^{51}$ erg s$^{-1}$ ($\sim 30 \times 10^{51}$ erg s$^{-1}$) at a post bounce time of 1.2 s (1.6 s). The simulations using Eqs. (1)–(2), the pure-OPE, Carenza formalism [23], give lower axion luminosities and also follow a different hierarchy with respect to the value of $\xi$ chosen. The highest axion luminosity arises with the assumption of $\xi = 1.3078$ and the lowest with $\xi = 0$. This difference can be explained by understanding the difference between the axion emissivity in [18] and pure OPE emissivity in [23] [i.e., our Eq. (1)]. Between these two formalisms, the difference arises in the numerical value of the coefficients in front of the $(g_{an}^2 + g_{ap}^2)/2$ term and the $(g_{an} + g_{ap})^2/4$ term. One of these differences is the positive (negative) sign in front the $\xi$, giving relatively more (less) emissivity for the nondegenerate case compared to the degenerate case for the Carenza (Brinkmann & Turner) formalism. The simulation using a spatial-varying $\xi$ expectantly shows a behavior in between these two limits. The luminosity transitions at $\sim 750$ ms postbounce from where the axion luminosity comes from a mix between the degenerate and nondegenerate conditions to where the axion emission seems to be dominated by $\xi = 0$ emission. As the PNS temperature drops during the PNS cooling, the degeneracy inside the core changes and the axion emission now takes place in a mainly degenerate medium, which explains this observed behavior.

### 2. 20 $M_\odot$

In order to explore the impact of axion emission on a well tested progenitor we choose the 20 $M_\odot$ progenitor from [38]. Unlike in our comparison to the Fischer *et al.* results, here we utilize the baseline SRO EOS from Ref. [36] although we maintain the same *ad hoc* artificial heating for these 1D simulations. We perform several simulations using a range of coupling constants, listed in Table I, following the formalism of Carenza *et al.* [23] presented in Sec. II B. This allows us to test the impact of the axion emission on the explosion mechanism and early cooling phase systematically by changing the coupling constant.

In Fig. 3, we plot the shock radius evolution versus the time postbounce. For all of the simulations, after the initial rapid shock expansion following bounce, the shock expands linearly up to 140 ms postbounce. Then, for a $f_a \geq 2 \times 10^8$ GeV, the shock continues to expand, hovers for $\sim 40$ ms at $\sim 300$ km, and then enters a runaway expansion leading to the final explosion at $\sim 250$ ms postbounce. This time marks the accretion of the silicon-oxygen shell interface. For $f_a = 1 \times 10^8$ GeV, the shock expansion occurs slightly earlier, entering into a runaway explosion at $\sim 240$ ms post bounce. The simulation with a $f_a = 5 \times 10^7$ enters a runaway explosion even earlier, beginning at $\sim 150$ ms postbounce. As we shall see below, the difference in the explosion times can be linked to the differences in the neutrino heating and ultimately the axion emission. Due to the *ad hoc* nature of the artificial neutrino heating, the precise shock expansion trajectory is a bit arbitrary. Varying the artificial heating factor can change this trajectory, but the results presented here and below regarding the hierarchy observed with the axion emission and its impact on the explosion does not depend on this precise factor.

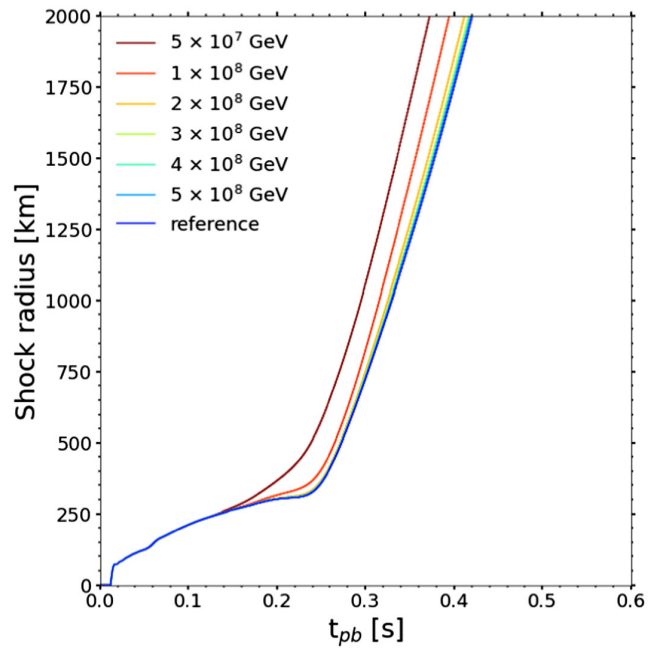

FIG. 3. Shock evolution with time in 1D for a 20 $M_\odot$ progenitor using artificial heating to trigger an explosion. The different colors correspond to different values of $f_a$ with the dark blue line being the reference simulation without axion emission.

In Fig. 4, we plot the axion luminosity (dashed lines), total neutrino luminosity (solid lines), and the total neutrino heating (dotted lines) as a function of time. For the total neutrino luminosity, in the first milliseconds following the core bounce, we observe a luminosity peak due to the strong electron neutrino release at bounce, the deleptonization burst. Then the luminosity decreases to $\sim 20 \times 10^{52}$ erg s$^{-1}$ and plateaus at this value up until $\sim 250$ ms postbounce, where it drops dramatically down to $\sim 9 \times 10^{52}$ erg s$^{-1}$. The timing of this drop corresponds to the accretion of the silicon-oxygen interface, and consequently to the onset of the explosion. The only outlier among the simulations is the one using $f_a = 5 \times 10^7$ GeV. In this case, we see a slightly higher neutrino luminosity during the plateau phase, peaking at 180 ms, followed by an earlier drop relative to the other simulations. This higher neutrino luminosity is tied to increased neutrino heating;

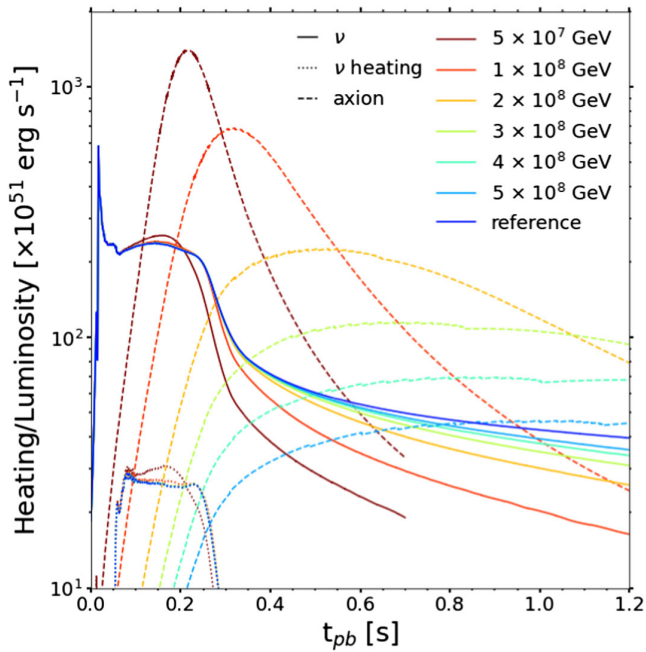

FIG. 4. Total neutrino (solid lines; measured at 500 km) and axion (dashed lines; integrated emission throughout the PNS) luminosity evolution with time. Additionally, we show the level of neutrino heating occurring during the accretion phase as dotted lines. The heating is defined to be the integral of the rate of neutrino energy deposition in the gain region. The latter is defined where the energy deposition rate is positive. The color code denotes the value of $f_a$, with blue being the reference simulation.

in fact the simulation with $f_a = 5 \times 10^7$ GeV has the highest neutrino heating in the first few 100s of milliseconds following bounce. As mentioned above, this is responsible for the earlier explosion time (see Fig. 3) for the simulation with this value of $f_a$. The simulation with $f_a = 1 \times 10^8$ GeV also shows slightly increased neutrino heating over the reference simulation, and again, a slightly earlier explosion time. Following the explosion and the steep drop in luminosity at ∼250 ms, the protoneutron stars enter the cooling phase and the neutrino luminosity now varies substantially with $f_a$.

To complement the evolution of the total neutrino luminosity evolution we show radial profiles of the local total neutrino luminosity in Fig. 5 for two select postbounce times, an early time (left panel; 0.182 s) and a late time (right panel; 0.982 s). The different colors correspond to different values of $f_a$ varying from dark red for the lower $f_a$ (higher coupling constant) to blue for the higher $f_a$ (lower coupling constant) and the reference run. Deep in the core the net flow of energy via neutrinos is actually inward, due to the diffusion away from the hot outer layers of the PNS. The bulk of the ultimate luminosity is generated outside of this hot compressed layer and has reached the asymptotic value by the edge of the plot. Here we also see the trends mentioned above, namely the divergent behavior at early times for low values of $f_a$ and the large spread of luminosities at later times. We will return to an explanation for these trends below following our discussion of the axion luminosity evolution.

For fixed thermodynamic conditions, in the free streaming limit, the axion luminosity, via, Eq. (1), scales with the coupling constant squared, or equivalently, $f_a^{-2}$. In the limit of very large $f_a$, or equivalently very weak coupling, the energy radiated in axions has no dynamical impact on the PNS, then we expect the timing profile of the axion emission to scale directly with the coupling constant squared. However, we are not in that regime for the values of $f_a$ chosen here, in fact it is our goal to capture any potential feedback from the axion emission on the dynamic and thermodynamic evolution of the PNS. Figure 4 shows the axion luminosity for our simulations with varying $f_a$ and Fig. 5 shows the radial profiles of the axion luminosity (dashed lines) for select times postbounce. It is important to note that the axion emission is mainly localized in the inner 15 km. The simulation with the smallest $f_a(= 5 \times 10^7$ GeV) initially has the largest axion luminosity. It peaks sharply at $\sim 1.5 \times 10^{54}$ erg s$^{-1}$ $\sim$ 200 ms after bounce. The energy loss due to axion emission in this case is strong enough to influence the early PNS evolution. In fact the axion emission not only cools the PNS core, but leads to a contraction of the PNS (via the loss of significant thermal pressure). This can be seen in the left panel of Fig. 6, in which we plot the density and temperature profiles for each simulation at three select postbounce times. Focusing on the $f_a = 5 \times 10^7$ GeV evolution compared to the reference evolution at 0.182 s postbounce, we can see that the axion-emission-induced contraction causes both the central density to increase and the outer mass shells of the PNS to heat up; both of these contribute to further increasing the axion emission. The axion emission nonlinearly impacts the PNS evolution. Eventually, the axions cool the PNS enough in the first 100s of milliseconds of the $f_a = 5 \times 10^7$ GeV simulation to slow down the axion emission. By 0.282 s (middle panel of Fig. 6), the temperature in the core has decreased enough such that the total axion luminosity is comparable to the $f_a = 1 \times 10^8$ GeV simulation (see Fig. 4). A similar trend is seen for increasingly higher values of $f_a$, but the time to the peak axion emission becomes longer as the axion emission is not as efficient at cooling the core. For the same reason, the subsequent decay of the axion emission is slower. For all the simulations we have performed, the time that the axion luminosity peaks and the rate at which it declines after this peak is directly related to the value of $f_a$ used.

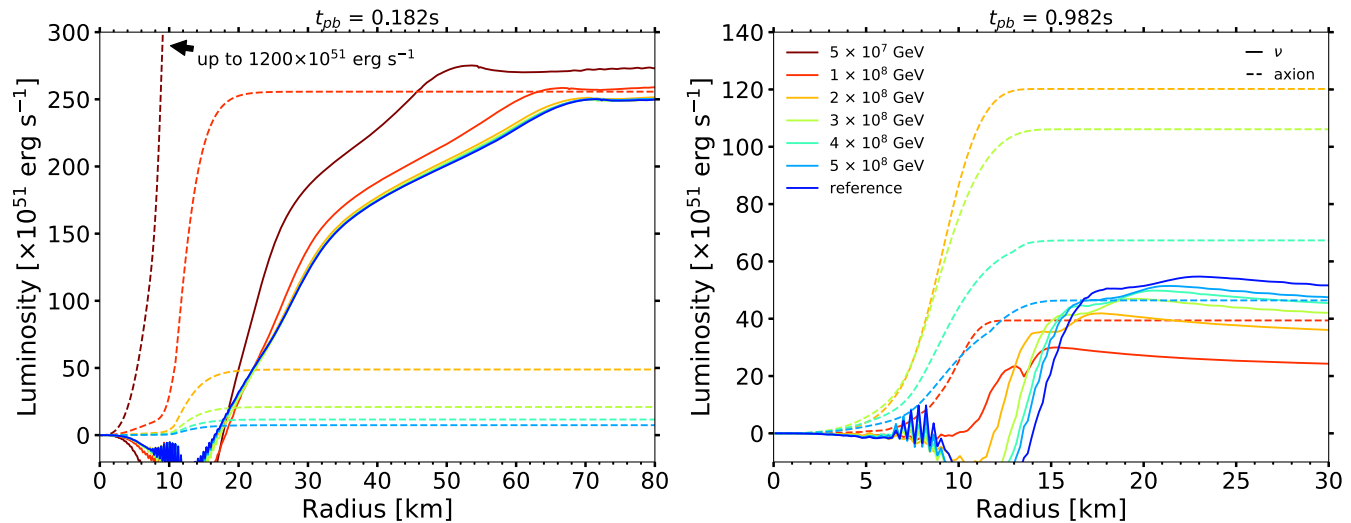

FIG. 5. Local neutrino (solid lines; all species summed together) and axion (dashed lines) luminosities at 0.182 s (left panel) and 0.982 s (right panel) postbounce for various coupling constants (denoted by colors as shown in the legend) for the 1D simulations of the 20 $M_\odot$ progenitor.

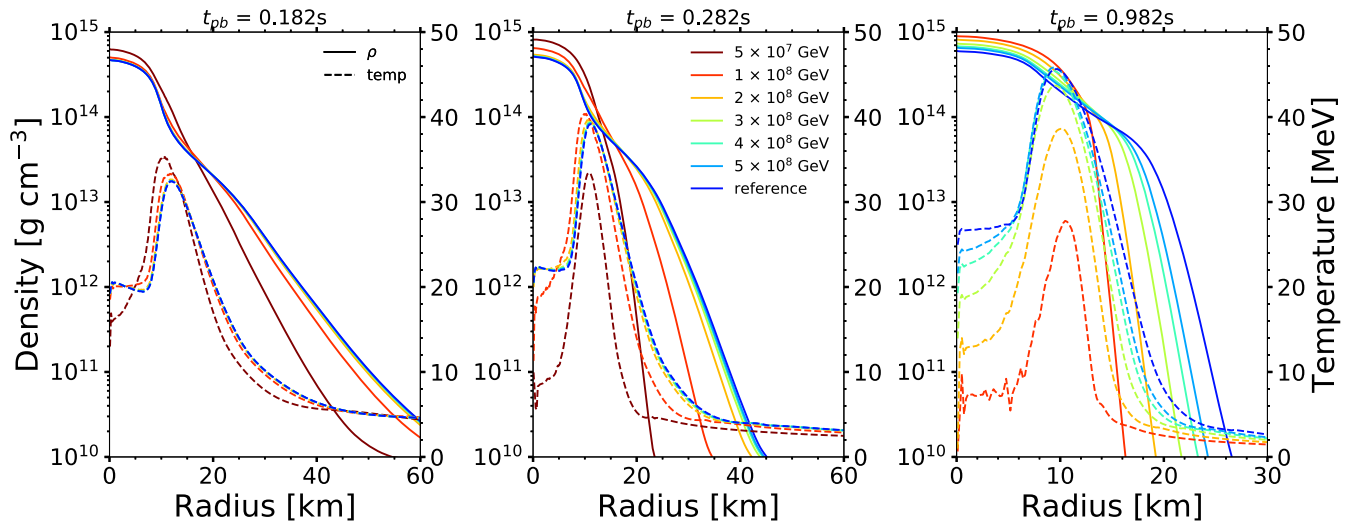

FIG. 6. Radial profiles of density (solid lines; left axis) and temperature (dashed line; right axis) for three different postbounce times of 0.182 s (left panel), 0.282 s (middle panel) and 0.982 s (right panel) for the 1D simulations of the 20 $M_\odot$ progenitor. For each time we show the profiles for each of the six values of $f_a$ and the reference run with no axions. Due to numerical issues the $f_a = 5 \times 10^7$ GeV simulations only progressed until 0.6 s and therefore is not shown in the last panel.

We note in the simulation with an extreme value of $f_a = 5 \times 10^7$ GeV that an incredible amount of energy in axions is emitted and by 0.6 s after bounce, the inner core has cooled enough to start causing problems in our hydrodynamic evolution, this limits our ability to simulate it past this point with our current software tools. We see similar signs of numerical problems with the higher values of $f_a$, for example for $f_a = 1 \times 10^8$ GeV at later times. In all cases, this is well after the explosion has been initiated.

With the aid of Figs. 5 and 6, we now proceed to discuss the feedback from the axion emission on the neutrino properties. We divide this into two parts, first focusing on the time near the onset of the explosion, then at a later time during the cooling phase.

Early in the evolution, at $\sim$0.182 s postbounce, the neutrino luminosities, the neutrino heating, and the shock radius have already been impacted by the axion loss for the simulation with $f_a = 5 \times 10^7$ GeV and to a somewhat lesser extent the simulation with $f_a = 1 \times 10^8$ GeV. Recall that both of these simulations had explosions before the reference simulation. At first thought this seems counter-intuitive: Why does increased cooling from axions give a higher neutrino luminosity? However, this can be understood by considering the impact that axion cooling has on the hydrodynamics. For these simulations, we see in the left panels of Figs. 5 and 6 that the PNS radii (defined as where the density is $10^{11}$ g cm$^{-3}$) and the radii where the outgoing neutrino emission is being generated are consistently lower due to the contraction induced by the axion emission in the core. Additionally, the matter temperatures in these regions are consistently higher than for the simulations with larger $f_a$. Quantitatively, we give the PNS radii and matter temperature at this radii for each simulation at postbounce times of 0.182 s, 0.282 s, and 0.982 s in Table II. This results in stronger neutrino emission and higher neutrino energies. The neutrino luminosities at 80 km are ∼10% larger for $f_a = 5 \times 10^7$ GeV and ∼3% larger for $f_a = 1 \times 10^8$ GeV when compared the to reference simulation (and all other values of $f_a$). The higher matter temperatures where the neutrinos decouple leads to higher neutrino average energies and increased neutrino heating (see Fig. 4) which in turn directly impacts the shock expansion (see Fig. 3) by triggering an earlier explosion time for these two values of $f_a$. Simulations with $f_a \geq 2 \times 10^8$ GeV show little differences in their neutrino luminosities at the onset of explosion. This is due to the lower previous axion loss and subsequent impact on the PNS evolution, which explains the similar explosion time and behavior for all these simulations.

TABLE II. PNS radii ($r_{\text{PNS}}$) and the matter temperatures at the corresponding PNS radius [$T(r_{\text{PNS}})$] for each 1D simulation using the 20 $M_\odot$ progenitor and the three select postbounce times shown in Fig. 6: 0.182s, 0.282s, and 0.982s.

| $f_a$ (GeV) | $t_{pb} = 0.182$s $r_{\text{PNS}}$ (km) | $T(r_{\text{PNS}})$ (MeV) | $t_{pb} = 0.282$s $r_{\text{PNS}}$ (km) | $T(r_{\text{PNS}})$ (MeV) | $t_{pb} = 0.982$s $r_{\text{PNS}}$ (km) | $T(r_{\text{PNS}})$ (MeV) |
|---|---|---|---|---|---|---|
| $5 \times 10^7$ | 38.4 | 5.82 | 21.2 | 4.71 | ... | ... |
| $1 \times 10^8$ | 47.3 | 5.18 | 29.9 | 4.49 | 15.5 | 4.05 |
| $2 \times 10^8$ | 49.7 | 5.02 | 35.8 | 4.26 | 18.1 | 3.83 |
| $3 \times 10^8$ | 50.1 | 5.00 | 37.0 | 4.22 | 20.4 | 3.67 |
| $4 \times 10^8$ | 50.2 | 4.99 | 37.4 | 4.20 | 21.7 | 3.63 |
| $5 \times 10^8$ | 50.3 | 4.99 | 37.6 | 4.20 | 22.6 | 3.58 |
| reference | 50.4 | 4.98 | 37.9 | 4.19 | 24.6 | 3.49 |

A faster cooling or contraction of the PNS has been shown previously as a positive factor on the CCSN explodability. While we indirectly impact the neutrinosphere properties by cooling the core, it can be much more efficient by cooling the hot mantle itself, for example via an increased heavy-lepton neutrino emission as the result of reduced neutral-current scattering opacity by, for example, strange quark contributions to the cross section [42], or from in-medium effects [48]. The EOS has also been shown to play a similar role. The thermal response of the EOS is sensitive to the nucleon effective mass [36,49]. High effective masses result in less thermal support in the cores and mantles of PNSs, this causes them to be more compact and the outer layers to be hotter. Completely analogous to what we see here, the more compact and hotter neutrinospheres result in core collapse evolutions more conducive to explosion [1,36].

We now focus on later times, ∼1 s after core bounce. The prior axion emission, for which the magnitude depends on the value of $f_a$, has cooled the PNS core to varying degrees. From Fig. 6, both the $f_a = 1 \times 10^8$ GeV and the $f_a = 2 \times 10^8$ GeV simulations have noticeably lower matter temperatures throughout the PNS and the PNSs themselves are much more compact. This is directly responsible (via lower temperatures but also through less surface area) for the significantly decreased neutrino luminosity emerging from these PNS compared to the reference model seen at this time in Fig. 4. For example, at ∼1 s the total neutrino luminosity calculated in the $f_a = 1 \times 10^8$ GeV simulation is 50% that of the reference model.

While we can see an impact of the axion emissivity in the first second of the collapse with these results for the values of $f_a$ explored, we would like to confirm with self-consistent simulations without the need for artificial heating. We therefore repeat many of these simulations using 2D, axisymmetric simulations. The results of these simulations will be described in the next section.

### B. 2D results

For our axisymmetric (2D) simulations, we use the same 20 $M_\odot$ progenitor, nuclear EOS, and set of neutrino interactions as for the 1D systematic study in Sec. III A 2, with the only difference being that we do not include the enhanced heating in the gain region as the multidimensional dynamics together with the self-consistent neutrino heating is sufficient to drive explosions if the conditions are favorable. We present the results of six axisymmetric simulations. We use five different values of $f_a$ ($1 \times 10^8$ GeV, $1.5 \times 10^8$ GeV, $2 \times 10^8$ GeV, $3 \times 10^8$ GeV, and $5 \times 10^8$ GeV) as well as a reference run where no axions are included. The goal is to test the impact of the axion emission in a self-consistent explosion while varying the coupling constant.

In Fig. 7, we show the shock radius evolution as a function of postbounce time from these six simulations. The color code is similar to the one used in Sec. III A 2, with the lowest value of $f_a$ ($1 \times 10^8$ GeV, and therefore the strongest coupling of the axions to the matter for the 2D simulations) being the dark red line and the reference simulation, with no axion emission, being the blue line.

For all the simulations, the shock initially expands in the same way with a shock radius increasing before reaching a peak value of ∼160 km at ∼100 ms postbounce, followed by a plateau phase lasting for ∼50 ms. For the simulation with the strongest axion-matter coupling, $f_a = 1 \times 10^8$ GeV, the shock recession is barely noticeable and the shock undergoes a runaway expansion beginning at a time of ∼250 ms postbounce. The simulations with higher values of $f_a$ show a longer shock stalling phase before the ultimate explosion. In general, the simulations reach the point of runaway shock expansion at times following a trend inversely proportional to $f_a$. The simulations with values of $f_a = 1.5 \times 10^8$ GeV and $f_a = 2 \times 10^8$ GeV show very similar behaviors and explosion times roughly 100 ms after the simulations with $f_a = 1 \times 10^8$ GeV. These explosions occur close to the time when the silicon-oxygen shell interface accretes through the shock. Associated with this event is a drop in the accretion rate that can often trigger an explosion if the simulation is "close". The remaining simulations seemingly do not fall in this category. Our simulation with a value of $f_a = 3 \times 10^8$ GeV explodes later with a shock radius evolution similar to that of the reference simulation. Our simulation with $f_a = 5 \times 10^8$ GeV explodes ∼150 ms later. We do not pursue the simulations significantly past the point of runaway shock expansion.

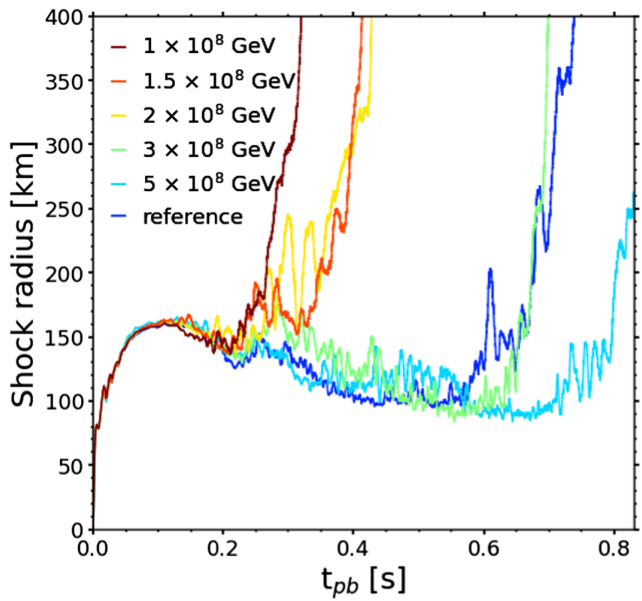

FIG. 7. Shock radius evolution for simulations assuming different values of $f_a$ in 2D axisymmetric simulations using a 20 $M_\odot$ progenitor.

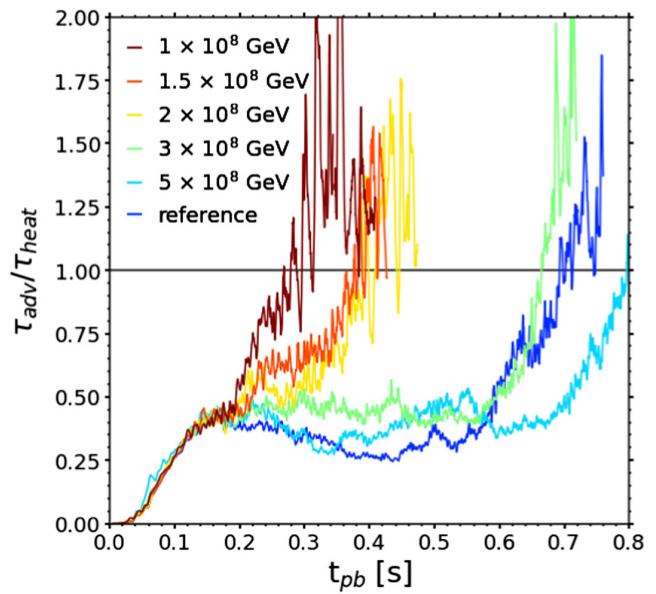

FIG. 8. Ratio of the advection timescale to the heating timescale as a function of postbounce time. This ratio is a measure of the closeness of a simulation to achieving an explosion. When the heating timescale is shorter than the advection timescale ($\tau_{\text{adv}}/\tau_{\text{heat}} > 1$) runaway shock expansion is expected.

The shock radius evolution is not the only, nor the best, measure we have on the closeness of a simulation to an explosion. To explore the differences between the simulations in more detail, we show the evolution of the ratio of the advection timescale over the heating timescale [50],

$$\frac{\tau_{\text{adv}}}{\tau_{\text{heat}}} = \frac{M_{\text{gain}}/\dot{M}}{|E_{\text{bind}}^{\text{gain}}|/\dot{Q}_{\text{heat}}}. \quad (5)$$

In this ratio, $M_{\text{gain}}$ is the mass of the gain region and $\dot{M}$ is the accretion rate through the shock, therefore their ratio represents an approximation for the advection time of mass through the gain region. $|E_{\text{bind}}^{\text{gain}}|$ is the binding energy of the matter in the gain region and $\dot{Q}_{\text{heat}}$ is the neutrino heating rate in the gain region; therefore their ratio is the time it takes to change the energy of the gain region by an amount equal to the binding energy. Quite generally, once the advection timescale exceeds the heating timescale, i.e., the ratio in Eq. (5) exceeds a value of 1, the matter in the gain region has time to absorb enough energy to unbind before it accretes out of the gain region. We therefore expect the shock to enter the runaway explosion phase near this time. Before the development of the explosion, this ratio serves as a convenient measure for the relative closeness of one simulation to an explosion when compared to a similar simulation. As shown Fig. 8, all the simulations reach $\tau_{\text{adv}}/\tau_{\text{heat}} \simeq 0.45$ at ∼150 ms postbounce. Consistent with the overall picture from Sec. III A 2, we find that this ratio of time scales increases fastest and is generally larger for the simulations with the lowest values of $f_a$. The simulation with $f_a = 1.5 \times 10^8$ GeV rapidly approaches a value of 1. For the simulations with $f_a = 1.5 \times 10^8$ GeV and $f_a = 2 \times 10^8$ GeV the rise is slower and the value of 1 is reached ∼100 ms later. The $f_a = 3 \times 10^8$ GeV simulation, and to a lesser extent the $f_a = 5 \times 10^8$ GeV simulation, maintains an advection-to-heating timescale ratio larger than the reference run, but not large enough at the time when the silicon-oxygen shell interface accretes in to launch an explosion. These simulations, as well as the reference simulation show a stalling phase of at least ∼500 ms.

All of these simulations, but especially the simulations which still have not exploded until late times, are subject to the impact of stochastic motions of the turbulent gain layer. This can have the effect of causing a wide range of explosion times for simulations differing only slightly in the initial conditions. For example, for this same progenitor model (albeit a different EOS) seeded with different initial perturbations, Refs. [28,48] showed a variation in explosion times of up to ∼200 ms. Therefore we caution against reading too closely into the precise ordering of explosion times in this work, especially for $f_a \geq 3 \times 10^8$ GeV. It is likely this stochastic effect is larger when assuming axisymmetry, as we have done here, than when performing simulations in full 3D.

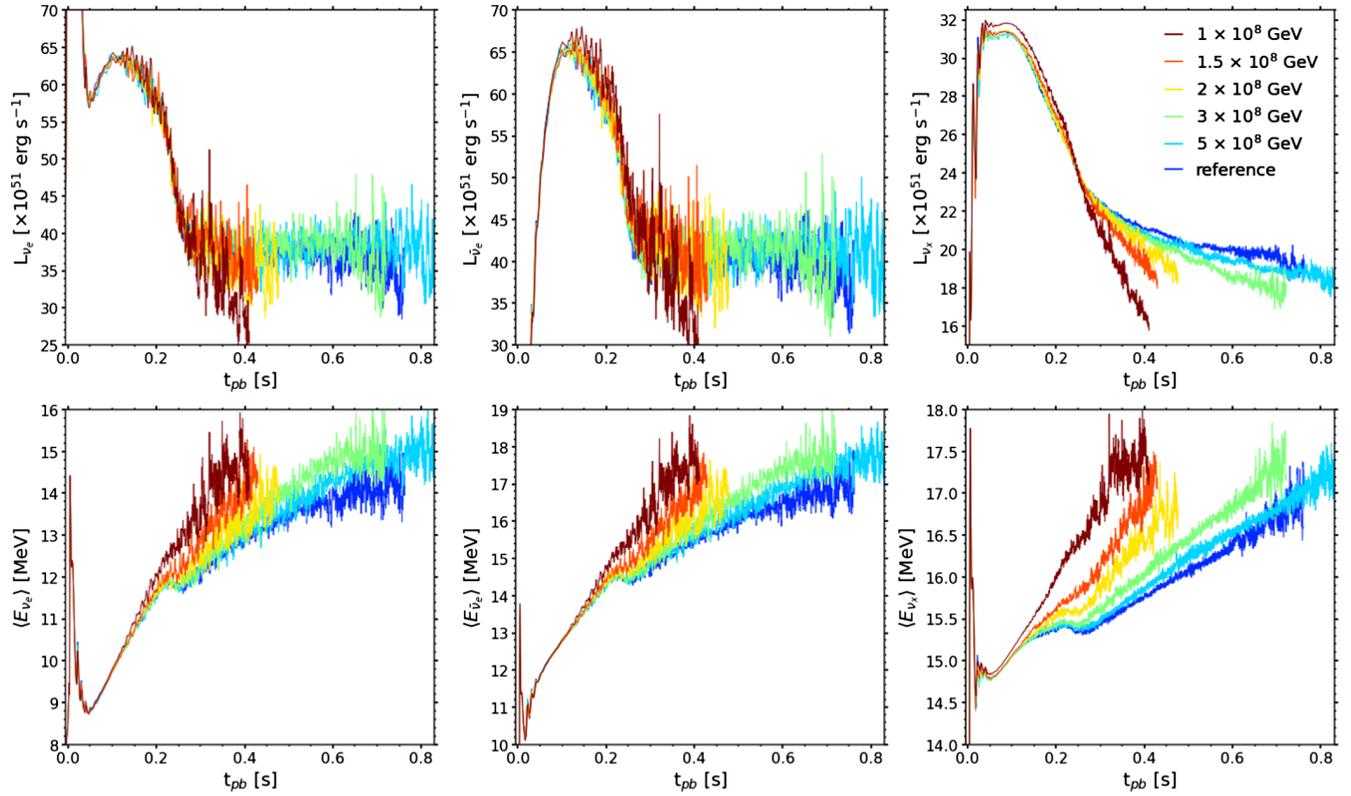

FIG. 9. Neutrino luminosities and mean energies for different values of $f_a$.

Nevertheless, as we saw with our parametrized models in 1D, but now in self-consistent multidimensional simulations, the increased axion-matter coupling is clearly driving earlier explosions. This is linked to the enhanced PNS cooling that is strongly impacted by the axion emission for values of $f_a \leq 2 \times 10^8$ GeV. The PNS contracts faster, releasing more gravitational binding energy and therefore impacting the temperature and the position of the neutrinosphere. This impacts the neutrino energies and therefore the neutrino heating in the matter behind the shock.

This impact can be seen in Fig. 9 where we show the individual neutrino luminosities (top row) and mean energy (bottom row) for the simulations with varying $f_a$. For the electron neutrino luminosities (upper left panel), there is initially no impact from the axion emission. The deleptonization neutrino burst lasts for ~50 ms at which point the neutrino luminosity dips to ~58 erg s$^{-1}$ before increasing and reaching a plateau value of ~64 erg s$^{-1}$ at ~150 ms postbounce. The plateau lasts until ~250 ms post bounce when the accretion rates drop because of the accretion of the silicon-oxygen shell interface and the neutrino luminosities decrease rapidly to reach a second plateau phase at

~40 erg s$^{-1}$, ~300 ms after bounce. This phase does not exist for the simulation with $f_a = 1 \times 10^8$ GeV as the accretion is slowed by the runaway shock expansion at this time. The other values of $f_a$ show a plateau phase with a duration equivalent to the shock recession phase before abruptly decreasing at the time of explosion. The electron antineutrino luminosities (upper middle panel) show a similar behavior with an initial plateau at ~65 erg s$^{-1}$, ~150 ms postbounce, a sharp decrease and a second plateau phase at ~300 ms postbounce reaching a value of ~40 erg s$^{-1}$. Once again, this second plateau lasts until the time of explosion, which is different for each simulation. Unlike the electron neutrinos, there is a slight hierarchy of plateau luminosity with $f_a$ with the $f_a = 1 \times 10^8$ GeV simulations having a slightly higher value. The heavy-lepton neutrino luminosities (upper right panel), emitted purely through neutral-current pair-processes and very sensitive to the temperature of the PNS [39], show a more varied evolution with $f_a$ before the runaway shock expansion. The luminosities reach a first plateau of ~31 erg s$^{-1}$ at ~50 ms postbounce which lasts for ~100 ms. The heavy-lepton luminosity for the $f_a = 1 \times 10^8$ GeV simulation is ~2–3% higher than the others at this point. This is because the heavy-lepton neutrinos are sourced from deeper in the PNS where the impact of the axion emission is more pronounced. The luminosities then abruptly decrease while the shock stalls and show a slower decrease once the shock begins to recede. We can see a trend of the heavy lepton luminosity declining faster as we increase the coupling constant (decrease $f_a$) and the explosions set in. There is a degeneracy here, where the variation in the decreasing heavy lepton neutrino luminosity can arise either from the drop in the accretion rate (like the electron type luminosities) or from the variation in the PNS structures due to the axion cooling.

Looking at all the neutrino luminosities, we note that the electron neutrino and antineutrino luminosities are less impacted by the axion emission. This is due to the charged-current interactions that electron type neutrinos and antineutrinos have with the matter. Relative to the heavy-lepton neutrinos, the electron type neutrinos remain coupled to the matter until much larger radii where they are less impacted by the changing PNS properties from the axion emission. The main difference between the different simulations is realized at the time of explosion where the electron neutrino and antineutrino luminosities decrease rapidly due to the decreased accretion of fresh material. This is not the case for the heavy lepton neutrinos. For example in the $f_a = 3 \times 10^8$ GeV and $f_a = 5 \times 10^8$ GeV simulations we see a decreased heavy-lepton neutrino luminosity prior to the explosion in these simulations even though their shock behavior and accretion rate is very similar to the reference. In these cases, the change in the temperature and the location of the neutrinosphere is impacted by the axion emission, but not enough to change the shock evolution behavior more than the stochasticity of the multidimensional turbulence does.

The neutrino mean energies are shown on the bottom row of Fig. 9. The electron neutrino mean energies (bottom left) show a peak at bounce, then decrease for ~50 ms before gradually increasing from ~9 MeV to ~12 MeV at ~200 ms after bounce. The mean energies then show either a short plateau phase for ~50 ms or, for the $f_a = 1 \times 10^8$ GeV simulation, a continuous increase as the explosion sets in. This plateau is expected and is related to the silicon-oxygen shell interface [43]. The rate of mean energy increase at this time shows a strong dependence on the value of $f_a$ as the axion emission impacts the PNS contraction rate and therefore the temperature at the neutrinosphere locations. We recall the behavior seen in Sec. III A 2, where the matter temperature at the radius where the density was $10^{11}$ g cm$^{-3}$ increased as the value $f_a$ decreased. The mean energy reaches ~15 MeV for all the simulations around the time of explosion. The electron antineutrino mean energies (lower middle panel) show a similar evolution as for the electron neutrinos. The mean energies after the first peak increase from ~10 MeV to ~14 MeV with short plateau at ~14.5 MeV at ~250 ms after bounce. The heavy-lepton neutrino mean energies (bottom right panel) are again evolving similarly although slightly slower. Over the course of the simulations the mean energy after the first peak evolves from a minimum around ~14.7 MeV to ~17.5 MeV. The steepness of the rise of the mean energies is correlated with the value of $f_a$; a lower $f_a$ in the simulation shows a faster increase of the neutrino mean energy. Here, the stronger axion emission for lower $f_a$ is driving contraction of the PNS and the increase in the emitted neutrino energies. For the $f_a = 1 \times 10^8$ GeV simulation, the mean energy is slightly higher than for the other simulations already ~100 ms after bounce. Relative to the electron type mean energies, we see a stronger variation in the heavy lepton mean energies among the different value of $f_a$. This is linked to the deeper neutrinospheres for the heavy lepton neutrinos.

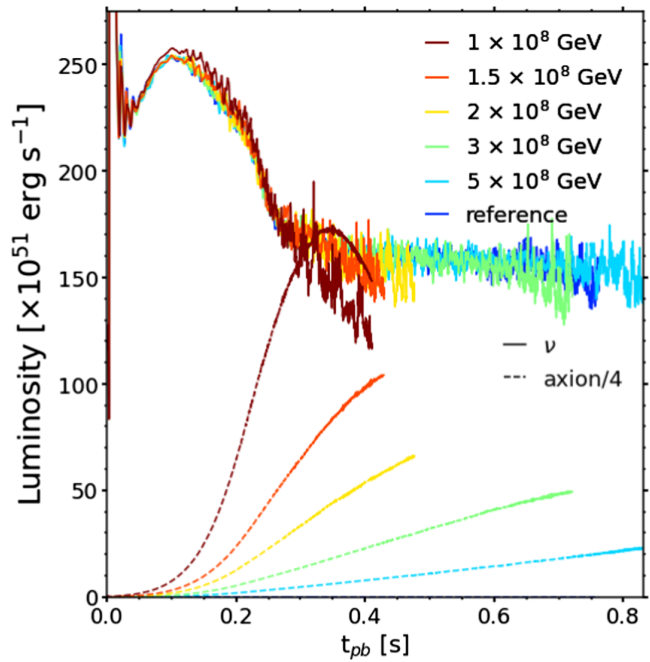

FIG. 10. Total neutrino (solid lines) and axion (dashed lines) luminosities versus the time for the five simulations with axion emission as well as the reference simulation. To add clarity to the figure we divide the axion luminosity by a factor of 4.

We show the evolution of the total neutrino luminosities as well as the axion luminosities with time in Fig. 10. The total neutrino luminosities are similar between all the simulations up to the time of explosion. The neutrino outburst of electron neutrinos occurs first, then the first plateau phase at ~250 erg s$^{-1}$, followed by a sharp decrease and a second plateau phase lasting until explosion. The axion luminosities that are plotted here are divided by 4 for clarity. This means for all of the values of $f_a \leq 3 \times 10^8$ GeV, the peak (or in some cases the final) axion luminosity is always higher than the neutrino luminosity at that time. The axion luminosity increases rapidly from ~100 ms after bounce up to the time of explosion. As in 1D, the axion emission eventually lowers the internal energy and temperature of the PNS enough to suppress further emission. One of the rough limits placed on the axion emissivity from SN 1987A is that the total energy emitted in axions should not be higher than the total neutrino energy emitted. While we do not simulate the entire cooling phase, we can nevertheless compute the total axion emission and total neutrino emission during the time of our simulations. The values are presented in Table III along with relative axion excess, i.e., $[\Sigma(L_a) - \Sigma(L_\nu)]/\Sigma(L_\nu)$. We want to highlight that the total luminosity sums are computed only during the simulated time. Of course, these will change while the PNS cools as will the relative fraction of the energy emitted between the neutrinos and the axions. The wide range of differences between neutrino and axion total luminosities shows the nonlinear impact of the coupling constant on the simulation. Recall from the 1D simulations, that a consequence of this nonlinearity is that it is not always the case that the simulation with the lowest $f_a$ always maintains a higher axion luminosity. We note the extreme nature of the $f_a = 1 \times 10^8$ GeV simulation where the total axion emission exceeds the total neutrino luminosity by ∼50% at the time when we stop the simulation. The excess itself and its implications for observations or limitations on the coupling constant will be discussed in Sec. IV.

TABLE III. Integrated luminosities for the axions and neutrinos along with the axion luminosity excess in percentages. These numbers are computed for the simulated time, which can be different among the simulations.

| $f_a$ (GeV) | $\Sigma(L_a)$ (erg) | $\Sigma(L_\nu)$ (erg) | $\frac{[\Sigma(L_a)-\Sigma(L_\nu)]}{\Sigma(L_\nu)}$ |
|---|---|---|---|
| $5 \times 10^8$ | $3.55 \times 10^{52}$ | $1.60 \times 10^{53}$ | $-77.9\%$ |
| $3 \times 10^8$ | $5.96 \times 10^{52}$ | $1.35 \times 10^{53}$ | $-56\%$ |
| $2 \times 10^8$ | $4.75 \times 10^{52}$ | $9.66 \times 10^{52}$ | $-50.8\%$ |
| $1.5 \times 10^8$ | $6.75 \times 10^{52}$ | $8.97 \times 10^{52}$ | $-24.7\%$ |
| $1 \times 10^8$ | $1.32 \times 10^{53}$ | $8.57 \times 10^{52}$ | $54.1\%$ |

## IV. DISCUSSION

### A. SN 1987A and axion emission

Axion emission from CCSNe has been the subject of many studies over the past few decades and has recently seen a resurgence, especially with a focus on placing limits on the different coupling constants between axions and matter. We implemented axion emission through the nucleon-nucleon bremsstrahlung, omitting other channels such as the emission though pion or photon decay. However, there is still the need for a discussion on the axion-nucleon coupling constant, and its limitations inferred by previous observations.

Some of the limits inferred for the axion mass and coupling constants come from the neutrino detection associated with the CCSNe explosion SN 1987A. These neutrinos were detected by some of the working neutrino detectors at the time, namely Kamiokande-II [51], IMB [52] and Baksan [53]. There are two main ways these neutrinos constrain the axion-matter coupling constant. First, the measurement of the total energy emitted in neutrinos places constraint on how much energy can be emitted in axions. Second, the duration of the neutrino signal relays to us how fast the PNS is cooling. This also limits the axion emission as axions act as an alternative cooling channel, reducing the cooling time and therefore the duration of the observed neutrino signal. The neutrinos associated with SN 1987A were detected over a time span of ∼12 s which means that the axion emission should not shorten the PNS cooling time so much as to place tension with this observation. Several limits have been inferred from these neutrino detections since SN 1987A. Most of these have been calculated using simple PNS cooling models or *a posteriori* emission calculations [18,20,21,24,54–56]. As mentioned before, only two studies [15,16] up to now have included axion emission in their full core-collapse simulations including the cooling phase and their goal was to study the impact of axion emission on the long term PNS cooling with modern CCSNe models. The values of the coupling constants for these works were chosen as "extreme", meaning that roughly equal energy is emitted in both neutrinos and axions over the entire cooling phase. This corresponds to values of $g_{ap} \sim 10^{-9}$ or $f_a \sim 5 \times 10^8$ GeV [15,16], which are on the lower end of the coupling strength values we consider. Our multidimensional results for these values of the coupling constants, for the one progenitor we have studied, agree with the conclusions of this prior work; i.e., the axion emission does not dramatically impact the explosion mechanism for these coupling strengths. We find that even more extreme values are needed to see an impact. However, we note the precise value of the coupling constant that gives equal total emission in neutrinos and axions is heavily dependent on the progenitor model, the EOS, and the variety of axion emission channels and formalisms included. Furthermore, we lack understanding of the evolution of the core of SN 1987A due to the low number of neutrino detections and our still maturing knowledge of the core-collapse process in general, especially the consequences of multidimensional effects on the late time evolution of the cooling and perhaps accreting PNS. Taking all of this together, we feel enabled to relax some constraints on the value of the coupling constants in order to study the multidimensional impact even though we concede some values perhaps may be too extreme.

The continued evolution of our models until later times, in order to self-consistently follow the nonlinear response of the axion emission, is needed to make direct comparisons with the SN 1987A emission, particularly in regards to the signal duration. This is left to future work. However, we can make some rough estimates. From Table III we can see that only the multidimensional simulation with $f_a = 1 \times 10^8$ GeV has an axion emission that exceeds the neutrino emission up until the end of the simulation. In the remaining simulations the neutrinos are still dominating the total emission at the point of explosion. Given the PNS baryonic mass at the end of the simulation ($\sim 1.8\ M_\odot$), and the binding energy of such a neutron star, we have already radiated $\sim 20$–$40\%$ of the binding energy in neutrinos in these multidimensional simulations and therefore the fiducial limit of comparable neutrino and axion emission is not far off.

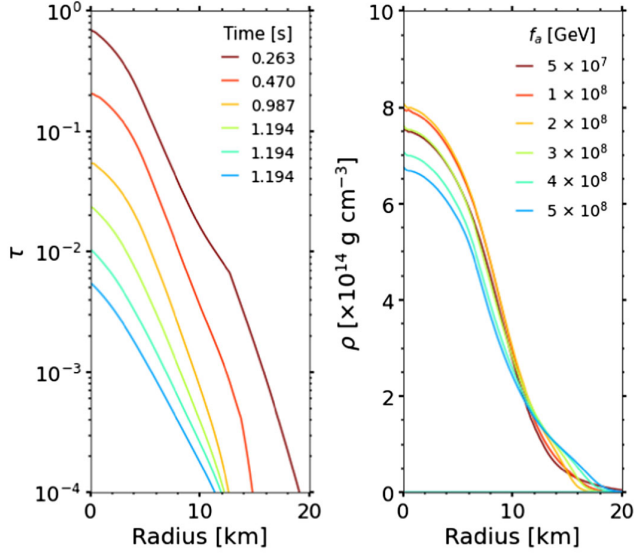

FIG. 11. Left: Grey optical depth as a function of radius for different values of $f_a$. Right: Density as function of radius for different values of $f_a$. All the optical depths are shown at their maximum, with the time indicated in the left panel. For all the runs with a value of $f_a \geq 3 \times 10^8$ GeV, the time of maximum is the end of the simulation as the axion emission is still not receding. For all the other values, the maximal optical depth is reached after the peak of axion emission, when the density is increasing in the core.

## B. Trapping vs free streaming

Another limitation in our simulations is the assumption of the free streaming regime for the axions. We therefore compute the optical depth for all our simulations, verifying this assumption. We highlight that the optical depth has been calculated on the axion emission implemented in our code, which is only based on the forward nucleon-nucleon bremsstrahlung channel. In order to do so, we took the formula described in the appendix of [23],

$$Q_a = \frac{T^4}{2\pi^2 \hbar^3 c^2 \rho} \int_0^\infty dx x^3\ e^{-x} \lambda_x^{-1}, \quad (6)$$

and assuming for simplicity a mean free path independent from the axion energy, we can invert this to obtain an estimate for the mean free path as

$$\lambda = \frac{3 T^4}{Q_a \rho \hbar^3 c^2 \pi^2}, \quad (7)$$

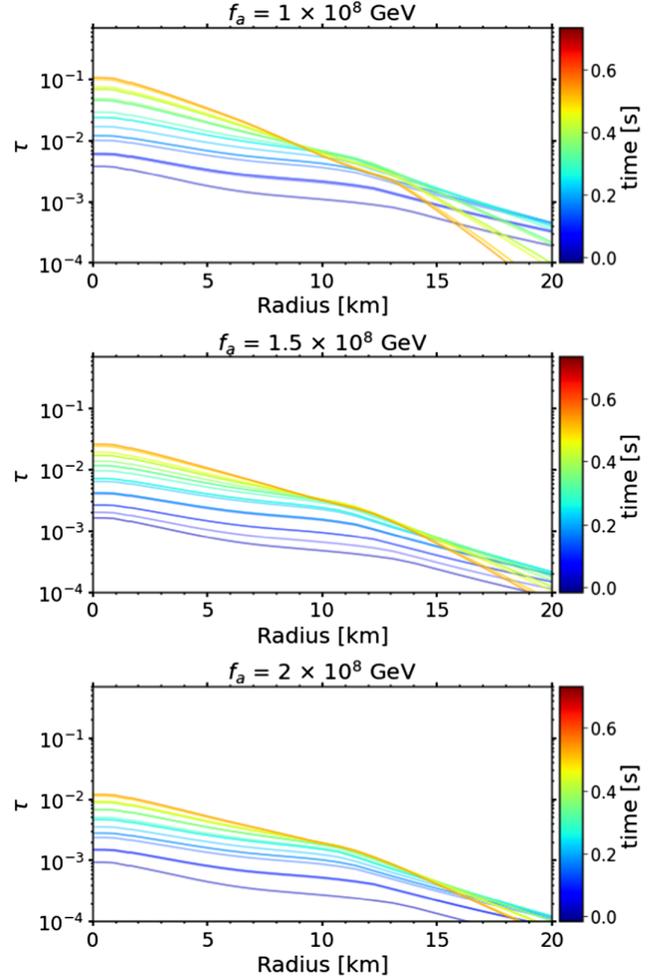

FIG. 12. Optical depth as a function of time and radius for different values of $f_a$. The color scheme represents the time evolution while each panel represents a simulation with a different value of $f_a$.

where $T$ and $\rho$ are the local matter temperature and density and $Q_a$ is the local specific axion emissivity as given by Eq. (1) and Eq. (2). As we are only interested in a rough estimate to assess our assumption of free streaming, the assumption of an energy independent mean free path is justified. Figure 11 shows the optical depth as a function of radius for all of the 1D, spherically symmetric simulations performed in Sec. III A 2 with different values of $f_a$. The radial-dependent grey optical depth is obtained by integrating the mean free path from the radius $r$ to infinity and therefore peaks at the origin. The displayed optical depth is for a postbounce time (shown in the legend in the left panel) when the central value of the optical depth is maximal throughout the simulation, generally close to the maximal axion emission. In the right panel, we show the density profile at the same postbounce time. The different colors correspond to different values of $f_a$. The maximum optical depth is reached for the simulation with $f_a = 5 \times 10^7$ GeV at $\sim 263$ ms postbounce. This maximum value corresponds to $\tau \approx 0.7$ and, therefore, the free streaming approximation is not completely valid for this value of $f_a$ and a more complete transport scheme should be considered. For all other simulations where the value of $f_a$ is larger, the optical depth remains low and confirms the assumption of free streaming for these simulations.

These results allowed us to perform our 2D simulations with values of $f_a \geq 1 \times 10^8$ GeV under the free-streaming assumption, treating the axion emission as a pure sink term for the internal energy. We also computed an angle averaged optical depth in order to verify the free-streaming assumption in 2D, shown in Fig. 12. Each of the subplots is showing an angle averaged $\tau$ depending on radius for different times postbounce. $f_a$ for each subplot varies from $1 \times 10^8$ GeV at the top to $2 \times 10^8$ GeV on the bottom. The maximum $\tau$ is reached for the simulation with $f_a = 1 \times 10^8$ GeV and is $\tau \simeq 10^{-1}$. This result, consistent with the 1D case, shows that axions are never fully trapped in our 2D simulations considering only the nucleon-nucleon bremsstrahlung channel. Using other channels of axion emission and/or scattering could potentially modify these results and break the free-streaming assumption but we focus on the nucleon-nucleon bremsstrahlung here.

## V. CONCLUSION

In this study, we looked at the impact of axion emission on the early stages of the development of an explosion in CCSNe simulations. We first reproduced the spherically symmetric simulations results of a previous study from Fischer *et al.* [16] using the formalism from [18] and two different degeneracy parameters $\xi$, then we added simulations using the formalism from [23] and three different degeneracy parameters $\xi$. In order to obtain an explosion in 1D and match the simulations of [16], we used an artificial neutrino heating scheme. This allowed us to test the impact of different formalisms, as well as the importance of the degeneracy parameter, for the nucleon-nucleon bremsstrahlung axion emission through the explosion phase. We then focused on 1D simulations using a well tested 20 $M_\odot$ progenitor and a range of values for $f_a$. As before, we invoke a parametrized explosion. These simulations showed an impact of the axion emission on the early explosion, creating a different behavior for the shock evolution when $f_a \leq 1 \times 10^8$ GeV. We attribute the divergent shock evolutions as a consequence of the axion emission. The enhanced cooling from the axion emission for small values of $f_a$ cooled and contracted the PNS core. The extra contraction of the outer layers of the PNS cause the thermodynamic properties at the neutrinosphere to change and enhance the neutrino heating.

Encouraged by these results we performed a select number of multidimensional simulations spanning the range of $1 \times 10^8$ GeV $\leq f_a \leq 5 \times 10^8$ GeV for this same progenitor. As was the case in 1D, but now with self-consistent neutrino heating in 2D, the variation in axion emission has an impact on the shock evolution with an explosion occurring earlier for lower values of $f_a \leq 2 \times 10^8$ GeV. We note that even though an earlier explosion is not realized in simulations performed with higher values of $f_a$ (weaker coupling), a quantitative impact is seen in the thermodynamics and hydrodynamics of some simulations through an increased value of the ratio $\tau_{\mathrm{adv}}/\tau_{\mathrm{heat}}$. Like in 1D, the early shock expansion is correlated with increased neutrino heating which follows from the specific PNS evolution in the early times post bounce, axion emission cools the core and contracts the PNS resulting in hotter outer layers where the neutrinos originate from. In this regime, axion emission aids the neutrino driven explosion mechanism. This is a novel result and counterintuitive at first glance, however the mechanism enabling the stronger explosions is well supported in the literature.

We also highlight that the heavy lepton neutrino properties are impacted by the axion emission much earlier than the electron type, often even before the onset of the explosion. This is due to the electron neutrino and anti-neutrino interactions being dominated by charged current weak interactions while heavy lepton neutrinos are dominated through neutral current weak interactions only. A consequence of these interactions is that the heavy-lepton neutrinos decouple much deeper in the PNS and therefore communicate the impact of the axion emission earlier.

In order to confirm the free-streaming assumption, we also postprocessed the simulations to calculate the mean-free path. These showed no value higher or equal to $\tau = 2/3$ [57], apart from the simulation in 1D with $f_a = 5 \times 10^7$ GeV. This implies that for the emission model used here, values of $f_a < 1 \times 10^8$ GeV, axions should not be treated through the free-streaming assumption but through an appropriate diffusion or transport scheme. We do note that this will be dependent on the progenitor model and EOS. We also have pointed out one of the possible limitations in our study linked to the total axion emission when compared to the total neutrino emission, which we discuss in Sec. IV A. The total axion emission should not be much higher than the total neutrino emission due to the confirmed detection of neutrinos from SN 1987A that are consistent, within the limits set by the low number of statistics, with an axion-free emission model but not one where the emission is dominated by axions [57]. In one of our simulations, the total axion emission is significantly higher than the total neutrino emission (up to ~92% for the time simulated) which would tend to rule out the associated values of $f_a$ ($f_a = 1 \times 10^8$ GeV). Furthermore, our simulations have been stopped short of the full evolution time and do not include the full PNS cooling after the explosion. During this time both axions and neutrinos will be emitted, but their evolutions are different depending on the temperature and density profiles, e.g., axions tend to decline faster. This must be done to fully characterize the ratio of emitted energies, and also to characterize the duration of the neutrino signal, which is also constrained by SN 1987A to some extent [15].

We also want to highlight the fact that our simulations only include the nucleon-nucleon bremsstrahlung channel of production for axions, which might not reflect the total axion emission when considering other channels such as photon or pion decay. A few studies [15,25] have shown that using channels including pions in their interaction could increase the axion production significantly. Their long time computation including axions shows that a value of $f_a \simeq 4 \times 10^8$ GeV does not produce a total axion emission higher than the total neutrino emission even including other interactions channels. We note, the inclusion of axion emission via pions, which has a much stronger temperature dependence, tends to shift the axion emission to early times [15]. This has consequences for the time duration of the neutrino signal, but, as we have shown here, should lead to a stronger hydrodynamic response.

The results coming from our 1D and 2D simulations show similar trends regarding the impact of axions on the final explosion. While 1D simulations are critical to assess the potential impact of axions or other exotic emission mechanisms, the need to artificially explode supernovae in 1D severely limits us from drawing conclusions on the explosion mechanism from 1D simulations alone.

Furthermore, the artificial explosion, and the 1D nature of the hydrodynamic response, can strongly impact the duration and strength of the neutrino and axion signals. Going forward, multidimensional simulations that self-consistently realize an explosion are key to obtaining accurate axion and neutrino signals.

## ACKNOWLEDGMENTS

We wish to thank Georg Raffelt for initial discussions surrounding the scope of this project. We also wish to thank Sean Couch for FLASH code development as well as Haakon Andresen, Pierluca Carenza, Andre da Silva Schneider, Oliver Eggenberger Andersen, and Shuai Zha for the fruitful discussions throughout the course of this work. The authors would like to acknowledge Vetenskapsrådet (the Swedish Research Council) for supporting this work under Award No. 2020-00452. The computations were enabled by resources provided by the Swedish National Infrastructure for Computing (SNIC) at NSC and PDC partially funded by the Swedish Research Council through Grant Agreement No. 2018-05973. We also acknowledge support from the ChETEC COST Action (CA16117), supported by COST (European Cooperation in Science and Technology).